\documentclass[twocolumn,tighten]{aastex631}

\received{January 5, 2023}
\revised{March 6, 2023}
\accepted{March 9, 2023}

\submitjournal{ApJL}

\shorttitle{Undiscovered UDGs in the Local Group}
\shortauthors{O.~Newton et al.}
\graphicspath{{./}{figures/}}
\usepackage[inline]{enumitem}
\usepackage{multirow}
\usepackage[slantedGreek]{newtxmath}
\usepackage{newtxtext}
\usepackage{tabularx}
\usepackage{catchfile}
\usepackage{etoolbox}
\usepackage{xifthen}
\usepackage{xspace}

\newcommand{\github}[1]{\href{https://github.com/#1}{\includegraphics[height=\fontcharht\font`\B]{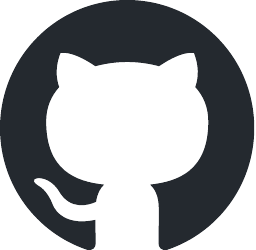} \nolinkurl{#1}}}

\newcommand{\targetlgmass}{\ensuremath{\Msun[8\times10^{12}]}}
\newcommand{\nMockSDSSHalo}{\ensuremath{{15,\!000}}}
\newcommand{\nMockSDSSSim}{\ensuremath{{30,\!000}}}

\newcommand{\NudgsInHestia}{one to four}

\newcommand{\fieldgalaxyresult}{\ensuremath{{52{\pm}7}}}
\newcommand{\maincombinedresult}{\ensuremath{{12{\pm}3}}}
\newcommand{\maincombinedsdssresult}{\ensuremath{{2^{+2}_{-1}}}}
\newcommand{\pZeroUDGinSDSS}{\percent{13.1}}

\newcommand{\eqnref}[1]{eq.~(\ref{#1})}
\newcommand{\figref}[1]{Fig.~\ref{#1}}
\newcommand{\figrefs}[2]{Figs~\ref{#1}~and~\ref{#2}}
\newcommand{\secref}[1]{Section~\ref{#1}}
\newcommand{\tabref}[1]{Table~\ref{#1}}
\newcommand{\exttab}[1]{table~\ensuremath{#1}}

\newcommand{\Arepo}{{\sc arepo}}
\newcommand{\Apostle}{{\sc APOSTLE}}
\newcommand{\Auriga}{{\sc auriga}}
\newcommand{\Hestia}{{\sc HESTIA}}
\newcommand{\RomulusC}{{\sc RomulusC}}

\newcommand{\AHF}{{\sc ahf}}
\newcommand{\astropy}{{\sc astropy}}
\newcommand{\numpy}{{\sc numpy}}
\newcommand{\pynbody}{{\sc pynbody}}
\newcommand{\python}{{\sc python}}
\newcommand{\scipy}{{\sc scipy}}
\newcommand{\matplotlib}{{\sc matplotlib}}

\newcommand{\DES}{\normalfont{DES}}

\newcommand{\SDSS}{\normalfont{SDSS}}

\newcommand{\code}[1]{\texttt{\detokenize{#1}}}
\newcommand{\percent}[1]{\ensuremath{#1\%}}
\newcolumntype{Y}{>{\centering\arraybackslash}X}

\newcommand{\band}[1]{{\textit{#1}--band}}
\newcommand{\LCDM}{{$\Lambdaup$CDM}}
\newcommand{\HI}{{H\thinspace\sc{i}}}

\newcommand{\Nfieldtot}{\ensuremath{N_{\rm field,\, tot}}}

\newcommand{\NUDGtot}{\ensuremath{N_{\rm UDG,\, tot}}}

\newcommand{\unit}[1]{\ensuremath{\mathrm{\,#1}}\xspace}
\newcommand{\unitlogicnospace}[2]{%
  \ifthenelse{\isempty{#1}}%
    {\unit{#2}}
    {\ensuremath{{{#1}\unit{#2}}}}
  }
\newcommand{\unitlogicspace}[2]{%
  \ifthenelse{\isempty{#1}}%
    {\unit{#2}}
    {\ensuremath{{{#1}\, \unit{#2}}}}
  }
\newcommand{\Msun}[1][]{\unitlogicspace{#1}{M_{\odot}}}
\newcommand{\sqdeg}[1][]{\unitlogicspace{#1}{deg^2}}

\newcommand{\asec}[1][]{\unitlogicspace{#1}{arcsec}}
\newcommand{\magn}[1][]{\unitlogicspace{#1}{mag}}
\newcommand{\magsqasec}[1][]{\unitlogicspace{#1}{\magn{}\, \asec{}^{-2}}}

\newcommand{\kpc}[1][]{\unitlogicspace{#1}{kpc}}
\newcommand{\pc}[1][]{\unitlogicspace{#1}{pc}}
\newcommand{\Mpc}[1][]{\unitlogicspace{#1}{Mpc}}
\newcommand{\Mpch}[1][]{\unitlogicspace{#1}{\mathnormal{h}^{-1}\, \Mpc{}}}

\newcommand{\variablelogicspace}[2]{%
  \ifthenelse{\isempty{#2}}%
    {\ensuremath{#1}}
    {\ensuremath{{{#1}{=}{#2}}}}
  }

\newcommand{\Mstar}[1][]{\variablelogicspace{M_{\ast}}{#1}}
\newcommand{\MV}[1][]{\variablelogicspace{M_{\rm V}}{#1}}
\newcommand{\mV}[1][]{\variablelogicspace{m_{\rm V}}{#1}}
\newcommand{\Rvir}[1][]{\variablelogicspace{R_{200}}{#1}}
\newcommand{\reff}[1][]{\variablelogicspace{R_{\rm e}}{#1}}
\newcommand{\ueff}[1][]{\variablelogicspace{\mu_{\rm e}}{#1}}
\newcommand{\z}[1][]{\variablelogicspace{z}{#1}}

\newcommand{\varwithbracket}[3]{%
  \ifthenelse{\isempty{#1}}%
    {\ensuremath{#3}}
    {\ensuremath{{{#3}\!\left(#2{#1}\right)}}}
  }
\newcommand{\MTot}[1][]{\varwithbracket{#1}{<}{M_{\rm LG}}}
\newcommand{\musb}[1][]{\varwithbracket{#1}{<}{\mu}}

\makeatletter
\patchcmd{\NAT@citex}
  {\@citea\NAT@hyper@{%
     \NAT@nmfmt{\NAT@nm}%
     \hyper@natlinkbreak{\NAT@aysep\NAT@spacechar}{\@citeb\@extra@b@citeb}%
     \NAT@date}}
  {\@citea\NAT@nmfmt{\NAT@nm}%
   \NAT@aysep\NAT@spacechar\NAT@hyper@{\NAT@date}}{}{}

\patchcmd{\NAT@citex}
  {\@citea\NAT@hyper@{%
     \NAT@nmfmt{\NAT@nm}%
     \hyper@natlinkbreak{\NAT@spacechar\NAT@@open\if*#1*\else#1\NAT@spacechar\fi}%
       {\@citeb\@extra@b@citeb}%
     \NAT@date}}
  {\@citea\NAT@nmfmt{\NAT@nm}%
   \NAT@spacechar\NAT@@open\if*#1*\else#1\NAT@spacechar\fi\NAT@hyper@{\NAT@date}}
  {}{}
\makeatother

\begin{document}

\title{The undiscovered ultra-diffuse galaxies of the Local Group}

\correspondingauthor{Oliver Newton}
\email{onewton@cft.edu.pl, adicintio@iac.es}

\author[0000-0002-2769-9507]{Oliver Newton}
\affiliation{Center for Theoretical Physics, Polish Academy of Sciences, al. Lotnik\'{o}w 32/46 Warsaw, Poland}
\affiliation{Univ. Lyon, Univ. Claude Bernard Lyon 1, CNRS, IP2I Lyon/IN2P3, IMR 5822, F-69622, Villeurbanne, France}

\author[0000-0002-9856-1943]{Arianna Di Cintio}
\affiliation{Universidad de La Laguna. Avda. Astrof\'{i}sico Fco. S\'{a}nchez, La Laguna, Tenerife, Spain}
\affiliation{Instituto de Astrof\'{i}sica de Canarias, Calle Via L\'{a}ctea s/n, E-38206 La Laguna, Tenerife, Spain}

\author[0000-0002-9990-4055]{Salvador Cardona--Barrero}
\affiliation{Universidad de La Laguna. Avda. Astrof\'{i}sico Fco. S\'{a}nchez, La Laguna, Tenerife, Spain}
\affiliation{Instituto de Astrof\'{i}sica de Canarias, Calle Via L\'{a}ctea s/n, E-38206 La Laguna, Tenerife, Spain}

\author[0000-0002-6406-0016]{Noam I. Libeskind}
\affiliation{Leibniz-Institut f\"{u}r Astrophysik Potsdam, An der Sternwarte 16, D-14482 Potsdam, Germany}
\affiliation{Univ. Lyon, Univ. Claude Bernard Lyon 1, CNRS, IP2I Lyon/IN2P3, IMR 5822, F-69622, Villeurbanne, France}

\author{Yehuda Hoffman}
\affiliation{Racah Institute of Physics, Hebrew University, Jerusalem, 91904, Israel}

\author[0000-0003-4066-8307]{Alexander Knebe}
\affiliation{Departamento de F\'isica Te\'{o}rica, M\'{o}dulo 15, Facultad de Ciencias, Universidad Aut\'{o}noma de Madrid, E-28049 Madrid, Spain}
\affiliation{Centro de Investigaci\'{o}n Avanzada en F\'isica Fundamental (CIAFF), Facultad de Ciencias, Universidad Aut\'{o}noma de Madrid, E-28049 Madrid, Spain}
\affiliation{International Centre for Radio Astronomy Research, University of Western Australia, 35 Stirling Highway, Crawley, Western Australia 6009, Australia}

\author[0000-0002-2307-2432]{Jenny G. Sorce}
\affiliation{Univ. Lille, CNRS, Centrale Lille, UMR 9189 CRIStAL, F-59000 Lille, France}
\affiliation{Univ. Paris-Saclay, CNRS, Institut d'Astrophysique Spatiale, F-91405, Orsay, France}
\affiliation{Leibniz-Institut f\"{u}r Astrophysik Potsdam, An der Sternwarte 16, 14482 Potsdam, Germany}

\author[0000-0001-6516-7459]{Matthias Steinmetz}
\affiliation{Leibniz-Institut f\"{u}r Astrophysik Potsdam, An der Sternwarte 16, 14482 Potsdam, Germany}

\author[0000-0002-5249-7018]{Elmo Tempel}
\affiliation{Tartu Observatory, University of Tartu, Observatooriumi 1, 61602 T\~oravere, Estonia}
\affiliation{Estonian Academy of Sciences, 10130 Kohtu 6, Tallinn, Estonia}



\begin{abstract}%
\noindent Ultra-diffuse galaxies~(UDGs) are attractive candidates to probe cosmological models and test theories of galaxy formation at low masses; however, they are difficult to detect because of their low surface brightness. In the Local Group a handful of UDGs have been found to date, most of which are satellites of the Milky Way and M31, and only two are isolated galaxies. It is unclear whether so few UDGs are expected. We address this by studying the population of UDGs formed in hydrodynamic constrained simulations of the Local Group from the \Hestia{} suite. For a Local Group with a total enclosed mass ${\MTot[{\Mpc[2.5]}]=\targetlgmass{}}$, we predict that there are \maincombinedresult{} isolated UDGs (\percent{68} confidence) with stellar masses $10^6 \leq{} \Mstar{}\, /\, \Msun{} < 10^9,$ and effective radii $\reff{} \geq \kpc[1.5],$ within \Mpc[2.5] of the Local Group, of which \maincombinedsdssresult{}~(\percent{68} confidence) are detectable in the footprint of the Sloan Digital Sky Survey~(\SDSS{}). Accounting for survey incompleteness, we find that almost the entire population of UDGs in the Local Group field would be observable in a future all-sky survey with a depth similar to the \SDSS{}, the Dark Energy Survey, or the Legacy Survey of Space and Time. Our results suggest that there is a population of UDGs in the Local Group awaiting discovery.%
\end{abstract}%



\section{Introduction}%
\label{sec:Introduction}%
Hierarchical models of galaxy formation predict the emergence of a large population of low-mass galaxies. Typically, they are dominated by sizable dark matter~(DM) components that make them useful as discerning probes of cosmological models. The most valuable galaxies for this purpose are those that contain little baryonic material, which is dispersed throughout a large volume. Such faint and extended galaxies were first characterized by \citet{sandage_studies_1984}, and a handful of additional systems were described subsequently \citep{impey_virgo_1988,thompson_dwarf_1993,jerjen_surface_2000,conselice_galaxy_2003,mieske_early-type_2007,de_rijcke_hubble_2009,penny_hubble_2009}. More recently, studies of this subpopulation of galaxies have been invigorated by the discovery of hundreds of systems in several different environments: within clusters of galaxies such as Coma, Virgo, and Fornax \citep{koda_approximately_2015,dokkum_forty-seven_2015,martinez-delgado_discovery_2016,roman_spatial_2017}; in galaxy groups \citep{trujillo_nearest_2017}; and in the field in between \citep[e.g.][]{leisman_almost_2017}. These extended objects have stellar masses and magnitudes typical of bright dwarf galaxies ($\Mstar[{\Msun[10^{6-9}]}]$ and $\MV{} < -8$, respectively); however, they are significantly larger, with sizes approaching those of massive galaxies such as the Milky Way. As a result they have very low surface brightness, usually between \ueff[24] and \magsqasec[28], earning them the sobriquet ``ultra-diffuse galaxies''~(UDGs).

The circumstances leading to the emergence of such diffuse galaxies are not understood fully and several scenarios have been proposed to explain their formation. These are divided broadly into two main categories:
\begin{enumerate*}[label=\roman*)]
    \item internal processes that drive stars toward the outer regions of the galaxy, as could happen in haloes with high spin \citep{amorisco_ultradiffuse_2016}, and during episodes of powerful stellar feedback \citep{di_cintio_nihao_2017,chan_origin_2018,cardona-barrero_metallicity_2023}; and
    \item the disturbance caused by external mechanisms such as stripping and tidal heating \citep{carleton_formation_2019,jiang_formation_2019,tremmel_formation_2020,benavides_quiescent_2021}, and galaxy mergers \citep{wright_formation_2021}.
\end{enumerate*}
A compelling test of these proposals requires a large sample of UDGs, the catalog of which has grown rapidly in recent years because of advances in instrumentation and observational techniques. However, UDGs remain challenging to detect so their census in the nearby universe is likely far from complete.

Similarly, the census of dwarf galaxies within the Local Group is also incomplete \citep{garrison-kimmel_elvis:_2014,newton_total_2018,nadler_modeling_2019,drlica-wagner_milky_2020,fattahi_tale_2020}. Using the \citet{di_cintio_nihao_2017} definition of UDGs, only eight Local Group galaxies satisfy the criteria: And~II, And~XIX, And~XXXII, Antlia~II, Crater~II, Sagittarius~dSph, WLM, and IC1613 \citep[][see also \citealp{mcconnachie_observed_2012} for observational data]{collins_kinematic_2013,kirby_dynamics_2014,torrealba_feeble_2016,caldwell_crater_2017}. Of these, six are satellites of the Milky Way and M31 and only IC1613 and WLM are found ``in the field'' of the Local Group, i.e. they are outside the virial radii of the Milky Way and M31, which we take to be $230$ and \kpc[275], respectively. It is unclear whether the dearth of UDGs in the field of the Local Group arises primarily from environmental influences that prevent most galaxies from becoming UDGs, or if observational limitations are the main obstacle impeding their detection. Indeed, if such a UDG population exists it would be partly obscured by the foreground of Milky Way stars and the background of other galaxies, making it difficult to detect with current instruments. Therefore, in this \textit{Letter} we use high-resolution simulations to quantify the number of UDGs that we expect to find in the field within \Mpc[2.5] of the Local Group, and study their potential detectability in current and forthcoming surveys.%
\section{Methodology}%
\label{sec:Methods}%
To estimate the size and properties of the population of UDGs in the field of the Local Group we require simulations that self-consistently model the formation and evolution of galaxies in this environment. The \Hestia{} suite does this \citep{libeskind_hestia_2020}, and consists of $13$ zoom-in simulations of Local Group analogs that were run with the \Arepo{} moving mesh code \citep{springel_e_2010} and the \Auriga{} galaxy formation model \citep{grand_auriga_2017}. Using estimates of the peculiar velocity field derived from observations \citep{tully_cosmicflows-2_2013}, the initial conditions are constrained to reproduce the major gravitational sources in the neighborhood of the Local Group. Consequently, at \z[0] the Local Group analogs are embedded in large-scale structure that is consistent with the observations when assuming the $\Lambdaup+$cold dark matter (\LCDM{}) cosmological model \citep[see e.g.][]{hoffman_constrained_1991,doumler_reconstructing_2013,sorce_cosmicflows_2016}.

The Local Group analogs are simulated at ``low'' and ``intermediate'' resolution in a \citet{planck_collaboration_planck_2014} cosmology. Three were resimulated at higher resolution using ${\sim}200{\rm M}$ DM particles in a high-resolution region consisting of two overlapping spherical volumes with radii of \Mpch[2.5], each centered on the Milky Way and M31 analogs at \z[0]. The spatial resolution achieved is \pc[177], and the effective masses of the DM and gas particles are $M_{\rm DM} = \Msun[{2\times10^5}]$ and $M_{\rm gas} = \Msun[{2.2\times10^4}]$, respectively. The simulations are labeled \code{09_18}, \code{17_11}, and \code{37_11}, after the initial seed used, and their physical properties can be found in \citet[][\exttab{1}]{libeskind_hestia_2020}. We use the Amiga Halo Finder (\AHF{}) algorithm \citep{gill_evolution_2004,knollmann_ahf_2009} to identify and characterize gravitationally bound structures in the simulations.

In the \Auriga{} model the star particles represent simple stellar populations of a given age, mass, and metallicity. Upon creation they are initialized using a \citet{chabrier_galactic_2003} initial mass function, and the mass that is subsequently lost due to stellar evolution is calculated using the yield curves from \citet{portinari_galactic_1998} and \citet{karakas_updated_2010}. The photometric properties of each star particle are computed using the \citet{bruzual_stellar_2003} stellar population synthesis model while neglecting the effects of dust attenuation. Consequently, the stellar populations may be as much as \magn[0.75] brighter in the \textit{V--} and \band{r}s that are of interest in this \textit{Letter} than they would be if dust attenuation was accounted for. However, we note that the isolated, low-mass galaxies in \Hestia{} experience minimal star formation activity at \z[0], so we expect that including dust attenuation would produce only a small effect.

\subsection{UDG selection criteria}
\label{sec:Methods:UDG_selection_criteria}
\begin{figure}%
    \centering%
	\includegraphics[width=\columnwidth]{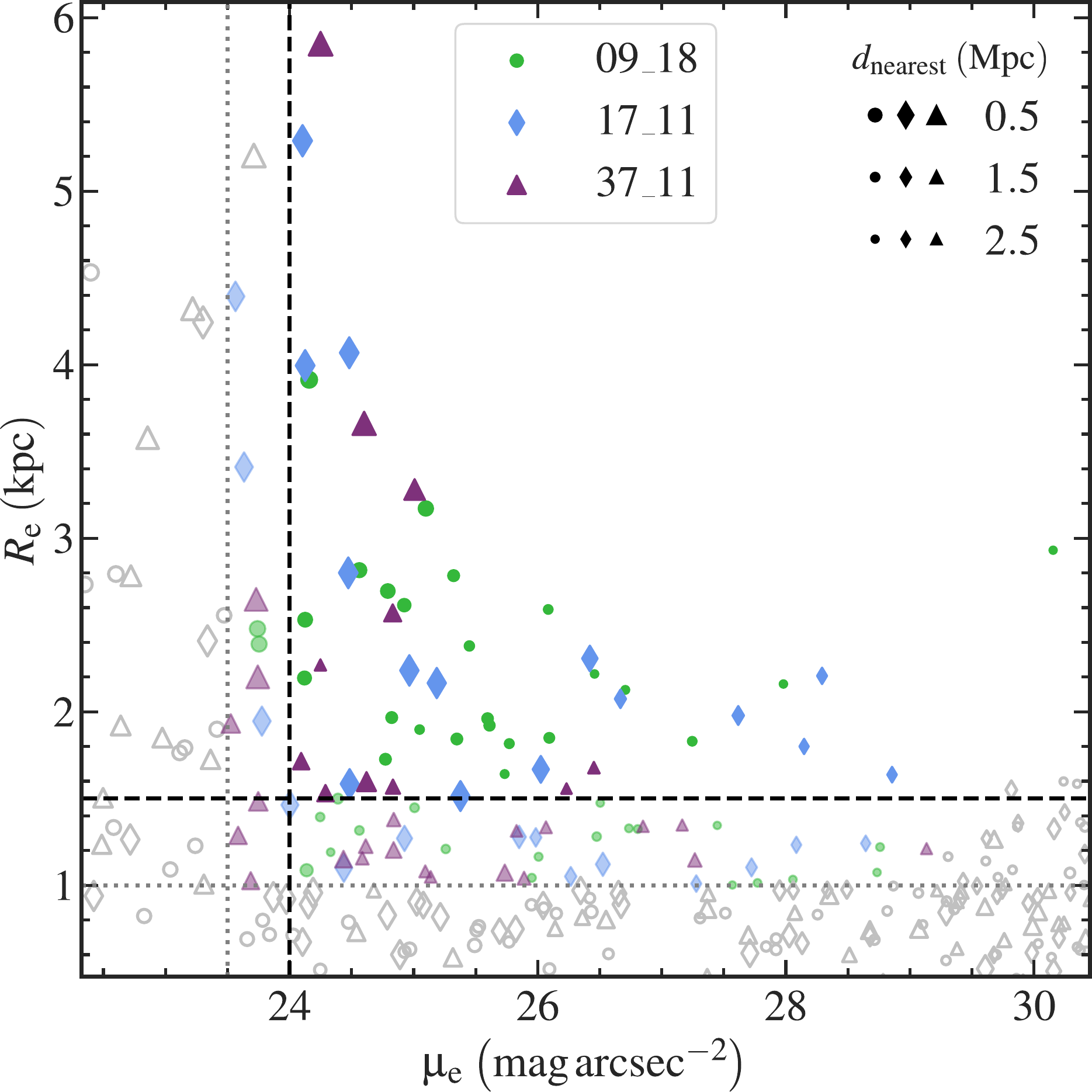}%
	\caption{The two-dimensional effective radius, \reff{}, as a function of the \band{r} effective surface brightness, \ueff{}, of the field UDGs in the three high-resolution \Hestia{} simulations. The size of each marker is inversely proportional to the distance to the nearest host galaxy. The dashed lines show two of our selection criteria applied to the field haloes in the simulations. The galaxies that satisfy all of the selection criteria described in \secref{sec:Methods:UDG_selection_criteria} are plotted with filled symbols, while unfilled symbols show the rest of the field galaxies. The faint filled symbols show galaxies that satisfy less stringent selection criteria (dotted lines) that are often used in the literature.
	}%
	\label{fig:Methods:re_vs_mu}%
\end{figure}%
The UDGs we study here are drawn from the population of Local Group field haloes in each high-resolution simulation. They are located within \Mpc[2.5] of the center of the Local Group at \z[0] and are outside \Rvir{} (the radius of the sphere enclosing a mean matter density of $\rho\!\left(< \Rvir{}\right) = 200 \times \rho_{\rm crit}$, where $\rho_{\rm crit}$ is the critical density for closure) of all haloes that are at least as massive as the Milky Way analog. We select  UDGs from the field haloes by applying criteria similar to  those described in \citet{di_cintio_nihao_2017}:
\begin{enumerate*}[label=\roman*)]
    \item the candidate has a total stellar mass, $\Mstar{} \leq \Msun[10^9]$;
    \item it has a two-dimensional effective radius, which contains half of the total luminosity of the system,
    ${\reff{} \geq \kpc[1.5]}$; and
    \item it has effective surface brightness,
    $\ueff{}{=}\musb[\reff{}] \geq \magsqasec[24]$.
\end{enumerate*}
Both \reff{} and \ueff{} depend on the luminosity of the galaxy, which we compute in the \band{r} while ignoring the effects of dust attenuation. We calculate these values by orienting the galaxy so that the gas disk is face-on to the observer and project the star particles into the plane of the disk. When a galaxy has no identifiable gas disk we take the simulation \textit{z}-axis to be normal to the disk plane. To minimize the effects of the limited simulation resolution we also require that each UDG has at least $50$~star particles. This is equivalent to imposing an effective minimum stellar mass of approximately $\log_{10}\left(\Mstar{}\, /\, \Msun{}\right)=6.05^{+0.09}_{-0.10}$ (\percent{68} scatter).

In \figref{fig:Methods:re_vs_mu}, we show two of the key selection criteria applied to the simulated field galaxies. The filled symbols show UDGs that satisfy the modified \citeauthor{di_cintio_nihao_2017} criteria described above, and have stellar masses in the range $\Mstar[{\Msun[{\left[10^6,\, 10^9\right]}]}]$. Larger markers indicate galaxies that are closer to one of the hosts. Generally, the largest UDGs are found in close proximity to the Milky Way and M31 analogs. There are $24$, $15$, and $11$ UDGs in the fields of the \code{09_18}, \code{17_11}, and \code{37_11} simulations, respectively. A detailed analysis of their formation histories will be conducted in a companion paper (S.~Cardona--Barrero et al. 2023, in preparation).

\begin{figure}%
    \centering%
	\includegraphics[width=\columnwidth]{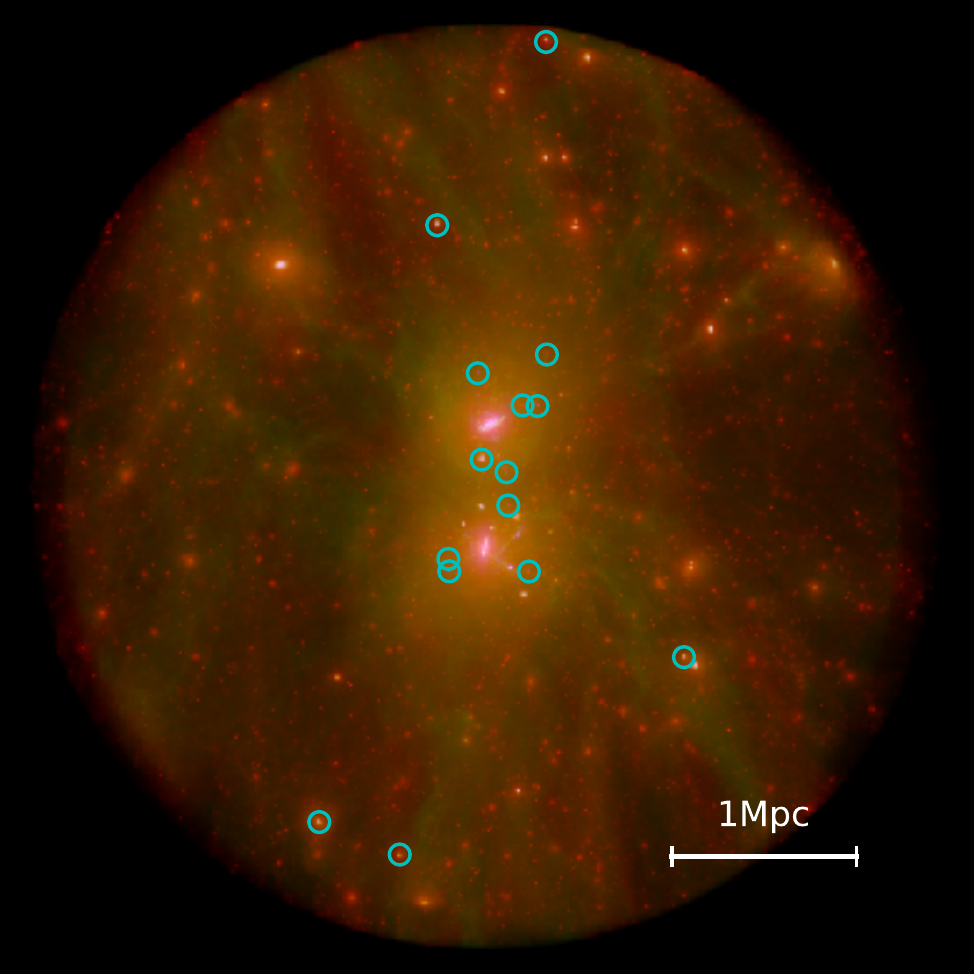}%
	\caption{The projected mass-weighted densities of DM (red), gas (green), and stars (white) within \Mpc[2.5] of the Local Group in the \code{17_11} simulation. The two bright galaxies at the center are the analogs of the Milky Way and M31 and we mark the projected positions of the UDGs with light-blue circles.}
\label{fig:Methods:LG_projection_UDGs_marked}%
\end{figure}%

In \figref{fig:Methods:LG_projection_UDGs_marked}, we show the distribution of the UDGs in one representative high-resolution simulation (\code{17_11}; chosen arbitrarily). This shows the projected DM, gas, and stellar density in a spherical region with a radius of \Mpc[2.5] centered on the midpoint of the Milky Way and M31 analogs. The distribution of UDGs throughout the volume is not uniform: at small distances from the center of the Local Group the UDGs cluster close to \Rvir{} of the Milky Way and M31 analogs, and in the other simulations they congregate nearer to the splashback radii of the hosts \citep[as defined in][]{diemer_flybys_2021}. However, at larger distances from the center of the Local Group the UDGs are affiliated preferentially with the large structures that compose the Local Group analog and the filaments and sheets that deliver matter to it.%
\section{Results}%
\label{sec:Results}%
The total number of field galaxies, \Nfieldtot{}, within \Mpc[2.5] scales with the total enclosed mass, \MTot[{\Mpc[2.5]}], of the Local Group \citep[see][]{fattahi_missing_2020}. This differs by a factor of $1.6$ between the least- and most-massive simulations and causes \Nfieldtot{} to vary between $50$~and~$79$~(see~\tabref{tab:Results:Mass_in_LGs}). The total number of UDGs, \NUDGtot{}, in each simulation varies between $11$ and $24$, and accounts for \percent{22}--\percent{30} of the total population of field galaxies in the stellar mass range $10^6 \leq \Mstar{}\, /\, \Msun{} \leq 10^9$. This is consistent with the results from the \RomulusC{} galaxy cluster simulation that shows that a large fraction of low-mass galaxies at \z[0] are UDGs \citep{tremmel_formation_2020}.

\begin{table}%
	\caption{The \z[0] properties of the three simulations. We provide the total enclosed mass, \MTot{}, the number of field galaxies, \Nfieldtot{}, and the number of UDGs, \NUDGtot{}, with $10^6 \leq \Mstar{}\, /\, \Msun{} \leq 10^9$ within \Mpc[2.5] of the midpoint of the primary haloes. Note the different observer position to that in \figref{fig:Results:Cumulative_distribution}.}%
	\label{tab:Results:Mass_in_LGs}%
	\begin{tabularx}{\columnwidth}{*{4}{Y}}%
	    \tableline
	    \multirow{2}{*}{Simulation} & \MTot{} & \multirow{2}{*}{\Nfieldtot{}} & \multirow{2}{*}{\NUDGtot{}} \\
	    & $\left(\Msun[10^{13}]\right)$ & &\\
	    \tableline
	    \code{09_18} & $1.23$ & $79$ & $24$ \\
	    \code{17_11} & $1.03$ & $58$ & $15$ \\
	    \code{37_11} & $0.77$ & $50$ & $11$ \\
	    \tableline
	\end{tabularx}%
\end{table}%

\begin{figure}%
    \centering%
	\includegraphics[width=\columnwidth]{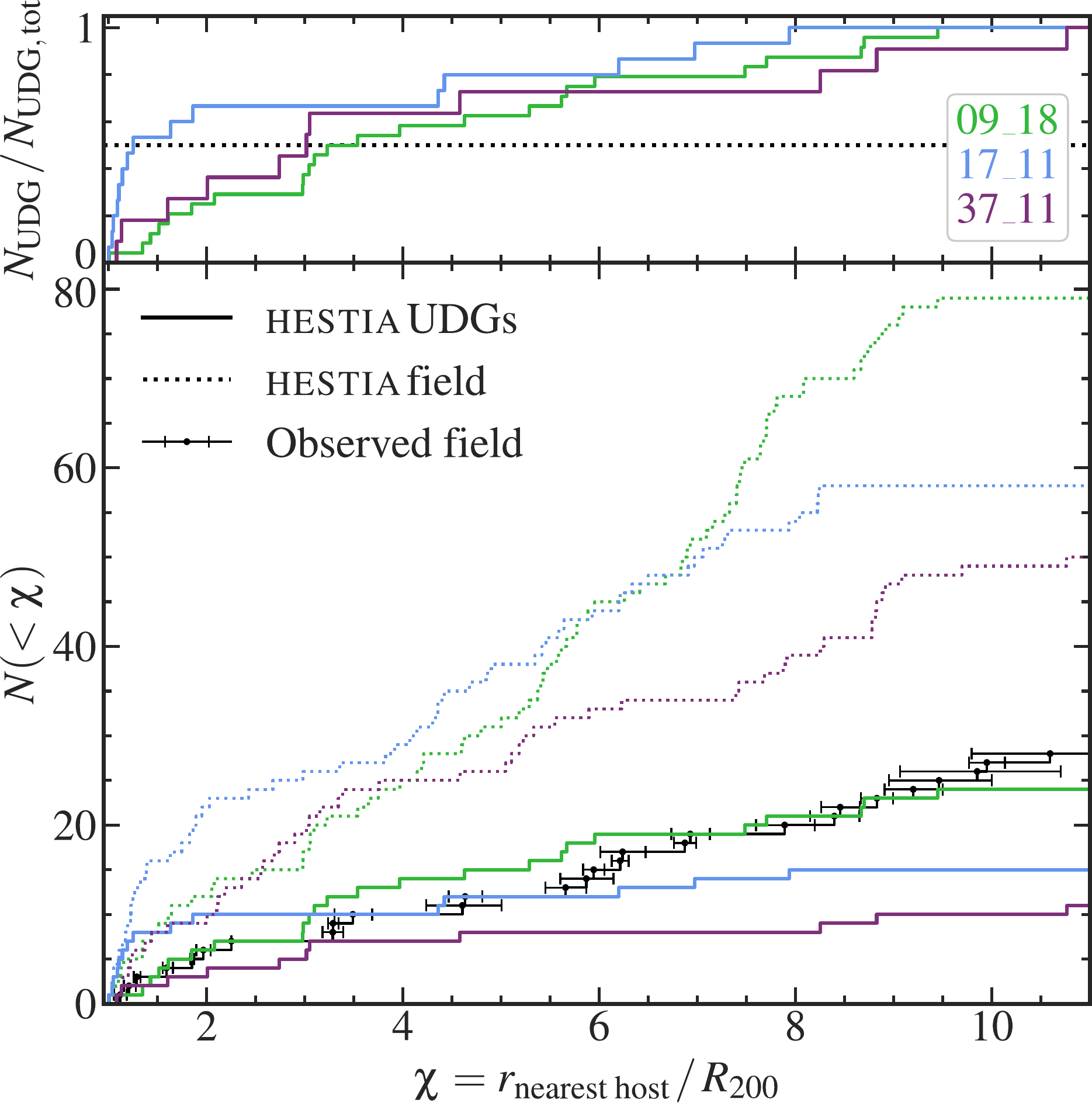}%
	\caption{\textit{Lower panel}: the whole-sky radial cumulative distributions of UDGs (solid lines) and all field galaxies (dotted lines) with stellar masses, ${10^6 \leq \Mstar{}\, /\, \Msun{} < 10^9}$, as a function of the distance to the nearest host galaxy, $r_{\rm nearest},$ normalized to the host galaxy's \Rvir{}. We overlay the incomplete census of observed field galaxies as points with error bars showing the \percent{68} distance uncertainties \citep[using data compiled by][]{mcconnachie_observed_2012}.
	\textit{Upper panel}: the cumulative distribution function of the UDGs in each \Hestia{} simulation. To aid the eye, we mark the \percent{50} threshold with a horizontal dotted line.
	}%
	\label{fig:Results:Cumulative_distribution}%
\end{figure}

In \figref{fig:Results:Cumulative_distribution}, we show the cumulative radial distributions of field galaxies and field UDGs in each high-resolution volume with respect to the nearest host galaxy analog at \z[0]. We also overlay the incomplete census of observed field galaxies in the stellar mass range described above. Their distances with respect to the Milky Way and M31 are calculated using the equatorial coordinates and distance moduli compiled in the most recently updated catalog of \citet{mcconnachie_observed_2012}. As we described in \secref{sec:Introduction}, we exclude galaxies that are within the virial radius of the Milky Way or M31. The observational data are limited by incomplete sky coverage and insufficient sensitivity to low-surface-brightness objects, which partly explains the discrepancy between these data and the number of field galaxies we identify in the simulations (see the bottom panel of \figref{fig:Results:Cumulative_distribution}). As we will discuss in \secref{sec:Results:Mock_sdss_LF}, we think that the \code{17_11} and \code{09_18} simulations could be $1.25-1.5$ times more massive than the Local Group. We expect that this would increase the number of field galaxies we identify in the simulations by a similar factor. Between \percent{62} and \percent{80} of the field UDGs in the simulations are found within \Mpc[1.5] of the center of the Local Group, and approximately half are between \Rvir{} and $3\times\Rvir{}$ of the host galaxies (see the upper panel of \figref{fig:Results:Cumulative_distribution}). The latter distance is consistent with the splashback radius defined in \citet{diemer_flybys_2021}. At larger radii, the UDGs are affiliated preferentially with the filaments and sheets that feed the growth of the Local Group (see also~\figref{fig:Methods:LG_projection_UDGs_marked}). This is in agreement with the results of \citet{fattahi_missing_2020}, who used the \Apostle{} simulations to show that most undiscovered dwarf galaxies should lie near the virial boundaries of the primary haloes.

\begin{figure*}%
    \centering%
	\includegraphics[width=\textwidth]{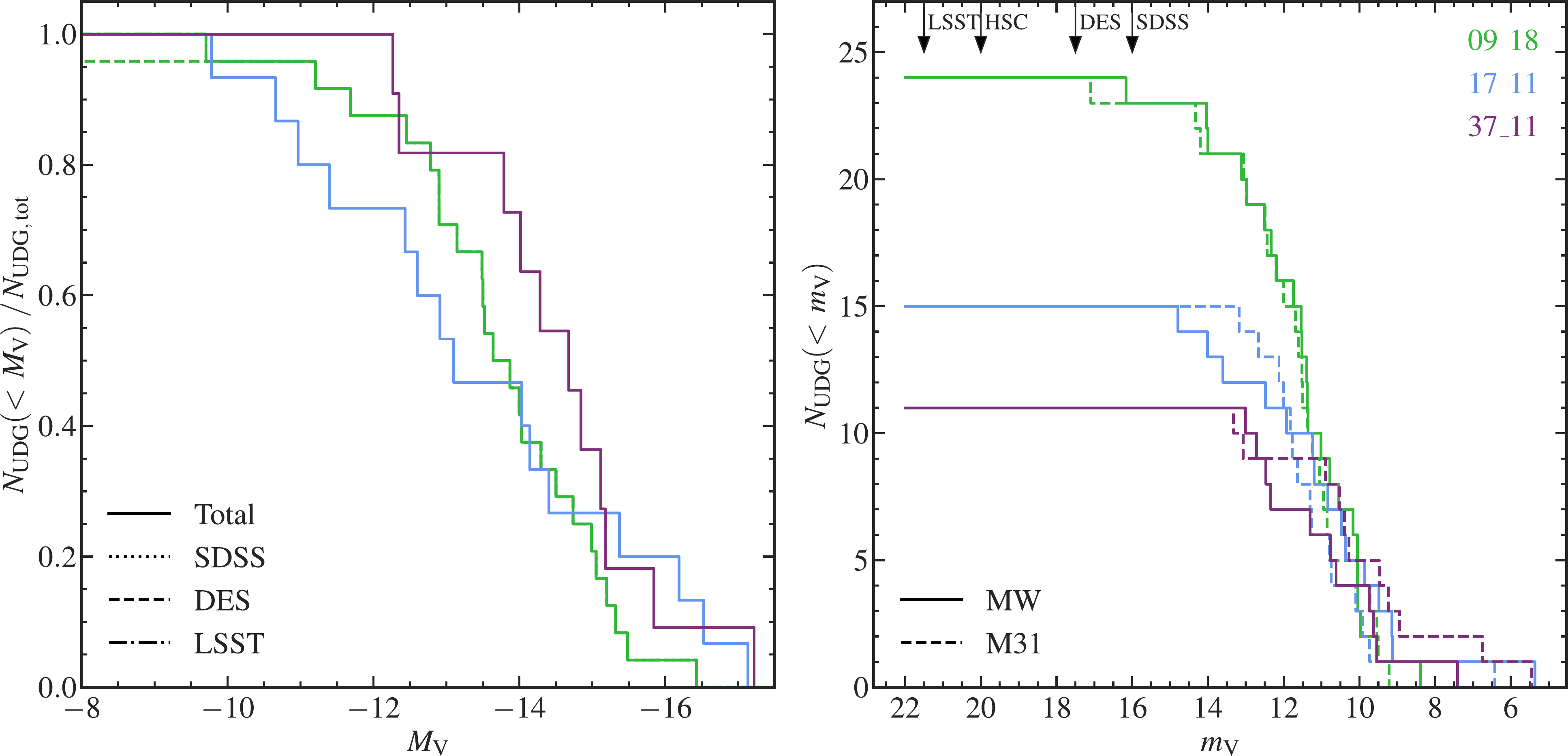}%
	\caption{The total luminosity functions of field UDGs within \Mpc[2.5] of each high-resolution Local Group simulation in the \Hestia{} suite.
        \textit{Left panel}: we estimate the absolute \band{V} magnitude luminosity functions as a fraction of the total number of UDGs, \NUDGtot{}, likely to be observed in whole-sky \SDSS{}-, DES-, and LSST-like surveys by applying a limiting surface-brightness cut in the \band{r} when assuming that the stars in each galaxy are individually resolved
        (\SDSS{}: \magsqasec[29], DES: \magsqasec[30], LSST: \magsqasec[31]).
        With the exception of the \SDSS{} and DES curves in the \code{09_18} simulation, the luminosity functions of all surveys overlap with the ``Total'' luminosity function.
        \textit{Right panel}: the cumulative number of field UDGs in the simulations as a function of apparent \band{V} magnitude, \mV{}. The solid and dashed lines show the luminosity functions measured by observers in the Milky Way and M31 analogs, respectively. The vertical arrows indicate the faintest dwarf galaxies that could be detected in several past and future surveys: \SDSS{} (\mV[16]), DES (\mV[17.5]), HSC (\mV[20]), and LSST (\mV[21.5]).%
	}%
	\label{fig:Results:combined_abs_app_vband_lf}%
\end{figure*}%

\subsection{Total luminosity functions}
\label{sec:Results:Total_LF}
The luminosity functions of the UDGs within \Mpc[2.5] of the center of each Local Group analog are shown in \figref{fig:Results:combined_abs_app_vband_lf}. All of the UDGs in the \Hestia{} simulations are as bright as the classical satellite galaxies of the Milky Way $\left(\MV{} < -8\right)$; however, they are much more diffuse and are close to the Milky Way and M31, which makes them difficult to detect in wide-area surveys using standard analysis techniques. To estimate how many UDGs could be observable in all-sky surveys with response functions similar to the Sloan Digital Sky Survey~\citep[\SDSS{};][]{blanton_sloan_2017}, the Dark Energy Survey~\citep[DES;][]{abbott_dark_2018}, and the Legacy Survey of Space and Time~\citep[LSST;][]{ivezic_lsst_2019}, we apply limiting surface brightness cuts in the \band{r} of $29,\, 30,$ and \magsqasec[31], respectively. We choose these with reference to the response function of the \SDSS{} determined by \citet{koposov_luminosity_2008}, and the design sensitivity of the LSST \citep{laine_lsst_2018}, both of which bracket the sensitivity of the DES. These limits apply to studies of resolved galaxies, and are several magnitudes deeper than the limits that are achievable in unresolved searches (V.~Belokurov 2022, private communication). Under these assumptions few of the \Hestia{} UDGs are too faint to be detected (\figref{fig:Results:combined_abs_app_vband_lf}, left panel). This simplified scenario suggests that almost the entire field UDG population is detectable in all-sky surveys with \SDSS{}-, DES-, and LSST-like surface-brightness limits. In practice, the surface-brightness limits depend on distance such that nearby very-low-surface-brightness galaxies are not observable \citep[see, e.g.][]{koposov_luminosity_2008}. We find that most of the field UDGs in \Hestia{} are close to the Milky Way and M31 analogs, so they could be difficult to detect; however, their surface brightnesses are high enough that they are detectable in the surveys.

In the right panel of \figref{fig:Results:combined_abs_app_vband_lf}, we plot the apparent \band{V} magnitude luminosity functions of the UDG populations. We generate luminosity functions for an observer located in the Milky Way and M31 analogs; however, there is little difference between them because the distributions of relative distances to the UDGs are similar. The magnitude limits of the surveys are marked with arrows and suggest that, on the basis of apparent magnitude alone, almost the entire UDG population is detectable in \SDSS{}-, DES-, and LSST-like surveys. Taken together, the panels in \figref{fig:Results:combined_abs_app_vband_lf} illustrate that UDGs in the Local Group should be detectable in all extant wide-area surveys. The most significant factors that could limit their detectability are likely to be their inclination with respect to the observer, which we discuss in \secref{sec:Results:Mock_sdss_LF}; and the obscuration of the sky by the Galactic disk, which we do not model here.

\subsection{Mock luminosity functions}
\label{sec:Results:Mock_sdss_LF}
The \Hestia{} simulations predict that UDGs exist in the field of the Local Group at \z[0] and that a fraction of them are potentially detectable by surveys such as the \SDSS{}. As very few field UDGs have been found to date, this suggests that several await discovery or that current models of galaxy formation do not accurately describe the physics at low masses. One test of this is to estimate how many UDGs we expect to find in the footprints of current surveys such as the \SDSS{} and whether they are, in principle, detectable using existing~data~sets.

To study this, we construct mock \SDSS{} observations of the population of field UDGs in the three simulations. This requires an understanding of the observational selection function of low-mass galaxies obtained by an algorithmic search of the survey data. Modern approaches to search for low-mass galaxies in wide-area surveys commonly adopt one, or both, of two complementary techniques:
\begin{enumerate*}[label=\roman*)]
    \item matched-filter searches that apply criteria to select samples of stars at a given distance and compare their spatial overdensity with the Galactic foreground \citep[e.g.][]{koposov_luminosity_2008,walsh_invisibles_2009}; and
    \item likelihood-based searches that model the properties of the stellar populations and incorporate observational uncertainties that are specific to the survey, such as the survey depth \citep[e.g.][]{bechtol_eight_2015,drlica-wagner_eight_2015}.
\end{enumerate*}
These are powerful techniques to search large areas of the sky efficiently but they are less sensitive than other methods to find spatially extended and low-surface-brightness galaxies. Approaches such as resolved star searches have been used very effectively to detect nebulous galaxies in small surveys like Hyper Suprime-Cam~\citep[HSC;][]{garling_case_2020}; however, they are impractical for wide-area sky searches. For this reason, in this study we use the selection function obtained by \citet{koposov_luminosity_2008}, who applied a matched-filter search to \SDSS{} data.

\citet{koposov_luminosity_2008} characterize the efficiency with which their algorithm detects galaxies with sizes up to \kpc[1] at distances as far as \Mpc[1] from the Sun. They do this using models of galaxies that are less spatially extended and closer than the field UDGs in the \Hestia{} simulations. Therefore, to apply their approach and estimate the \Hestia{} UDG detection efficiency in \SDSS{}, we extend to larger effective radii at greater distances from the Milky Way the relationships they calculated for the parameters in their matched-filter algorithms. This means that the detectability of the most distant galaxies could be overestimated because we do not account for star-galaxy confusion that most likely dominates the signal at large distances. Furthermore, we also disregard the effects of dust attenuation on the UDGs, and their possible obscuration by the Milky Way at Galactic latitudes $\left|b\right|\leq 10^\circ$, known as the Zone of Avoidance~(ZoA). Our results should therefore be interpreted as an upper bound on the detectability of UDGs in the \SDSS{} footprint when using this search algorithm. Using the analytic form provided by \citet{koposov_luminosity_2008}, the detection efficiencies, $\epsilon$, of the UDGs in the \SDSS{} are given by%
\begin{equation}
    \epsilon\!\left(\MV{},\, \mu\right) = G\!\left(\MV{} - M_{\rm V,\, lim}\right) G\!\left(\mu - \mu_{\rm lim}\right)\,,
    \label{eq:Results:epsilon}
\end{equation}
where
\begin{equation}
    G\!\left(x\right) = \frac{1}{\sqrt{2\pi}} \int^{\infty}_x \exp{-\frac{t^2}{2}}\, dt\,.
\end{equation}
We infer the limiting absolute \band{V} magnitude, $M_{\rm V,\, lim}$, and the limiting surface brightness, $\mu_{\rm lim}$, at distances greater than \Mpc[1] using a linear fit to the relationships in \citet[][fig.~12]{koposov_luminosity_2008}.

To generate a mock observation, we place an observer at the center of one of the primary haloes. We model the mock survey as a conical volume with an opening angle of \sqdeg[14,\!555], corresponding to the sky coverage of the \SDSS{}, and orient it so that its apex coincides with the observer. To account for the effects of the viewing angle, we assign each UDG a random orientation with respect to the observer and recalculate \reff{} and \ueff{}. UDGs that fail the selection criteria described in \secref{sec:Methods:UDG_selection_criteria} are discarded before the analysis proceeds. Using the relative distances of the UDGs with respect to the observer and the recomputed values of \ueff{}, we calculate $\epsilon$ using \eqnref{eq:Results:epsilon}. This represents the probability of detecting each UDG, and we use it to randomly select a set of UDGs that are detectable in the mock survey. As most galaxies have $\epsilon{\sim}1$ the effect of the random sampling is small.

\begin{figure}%
    \centering%
	\includegraphics[width=\columnwidth]{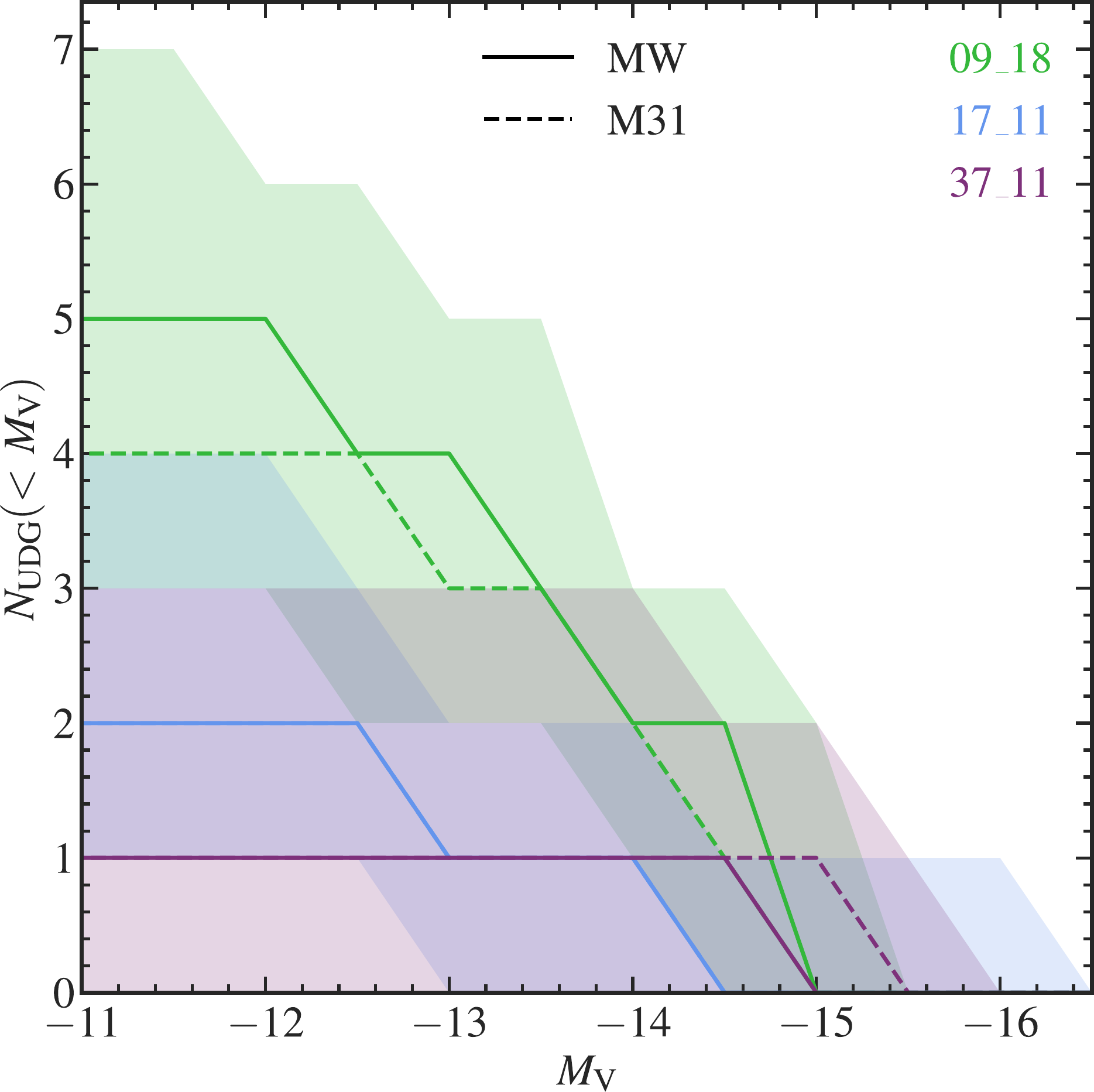}%
	\caption{Mock \SDSS{} observations of the population of field UDGs. The solid and dashed lines show the median predictions obtained by observers in the Milky Way and M31 analogs, respectively. The shaded regions represent the \percent{68} scatter in the Milky Way analog luminosity functions over \nMockSDSSHalo{} mock observations.}%
	\label{fig:Results:mock_sdss_lfs}%
\end{figure}%

We repeat this procedure for \nMockSDSSHalo{} pointings of the mock survey distributed evenly across the sky, and again for an observer in the second primary halo. We find that \nMockSDSSSim{} mock observations in each high-resolution simulation produces results that are well converged. Using these, we compute the medians and \percent{68} scatter of the field UDG luminosity functions that are detectable in \SDSS{} (see \figref{fig:Results:mock_sdss_lfs}). From this, in an \SDSS{}-like survey we find \NudgsInHestia{} UDGs within \Mpc[2.5] of the Milky Way analogs with \ueff{} brighter than \magsqasec[29] in the~\band{r}. Approximately \percent{44} of the simulated UDGs that are detectable in each mock \SDSS{} observation are misclassified as non-UDGs because of projection effects arising from their orientation with respect to the observer. Only five~per~cent of the mock observations contain at least one simulated UDG that is not detected at all. The projection effects impose the most significant limitation on the discoverability of UDGs in extant surveys. As we showed in \secref{sec:Results:Total_LF}, survey incompleteness has only a minimal effect on the number of UDGs that can be found.

As stated earlier, the total number of field galaxies depends strongly on \MTot{}, which is different in each \Hestia{} simulation. To account for this, we rescale the total mass of each simulation to ${\MTot[{\Mpc[2.5]}]=\targetlgmass{}}$ and adjust the number density of field galaxies according to the enclosed Local Group mass--galaxy number density relationship in \citet{fattahi_missing_2020}. Our choice of the mass enclosed within \Mpc[2.5] is motivated by current observational estimates of ${\MTot[{\Mpc[1]}]=\Msun[{\left[3,\,4.75\right]\times10^{12}}]}$ \citep{lemos_sum_2021,carlesi_estimation_2022,hartl_local_2022}. We use the mass profiles of the simulated Local Groups to extrapolate these values to an outer radius of \Mpc[2.5] and select the average mass. From this, we expect to find \fieldgalaxyresult{}~(\percent{68} confidence, CL) field galaxies with stellar masses $10^6 \leq M_\ast \, /\, {\rm M_\odot} < 10^9$ within \Mpc[2.5] of the center of the Local Group. Of these, approximately one-quarter (\maincombinedresult{}) are UDGs, and \maincombinedsdssresult{}~(\percent{68} CL) of them should be detectable in a reanalysis of the footprint of the \SDSS{}. Conducting a similar analysis for the DES and LSST using the \citet{koposov_luminosity_2008} response function with deeper surface-brightness and magnitude limits (by $1$ and \magn[2], for each respective survey), we expect that $3^{+3}_{-2}$ UDGs are detectable in the combined \SDSS{}+DES footprint, and $4\pm2$ will be detectable in the LSST (see \tabref{tab:Results:NUDGs_LG_mass_sel_criteria}). Using our selection criteria, no UDGs have been observed in the \SDSS{} footprint to date. Disregarding the effects of dust attenuation, we estimate that the chance that there are no field UDGs detectable in the \SDSS{} footprint is less than \pZeroUDGinSDSS{}. In \tabref{tab:Results:NUDGs_LG_mass_sel_criteria} we provide the predicted number of field galaxies and UDGs for different choices of \MTot{} and other UDG selection~criteria.

\begin{table*}%
	\centering%
	\caption{The total number of field galaxies, \Nfieldtot{}, UDGs, \NUDGtot{}, and the number of UDGs detectable in the \SDSS{} footprint, $N_{\rm UDG,\, SDSS}$, for combinations of \MTot[{\Mpc[2.5]}] and UDG selection criteria. Our fiducial choice is in bold.}%
	\label{tab:Results:NUDGs_LG_mass_sel_criteria}%
	\begin{tabularx}{\textwidth}{*{8}{Y}}%
    \tableline
\MTot[{\Mpc[2.5]}] & $R_{\rm e}$ & $\mu_e$ & \multirow{2}{*}{\Nfieldtot{}} & \multicolumn{4}{c}{$N_{\rm UDG}$} \\\cline{5-8}
$\left(\Msun[10^{12}]\right)$ & $\left(\kpc{}\right)$ & $\left(\magsqasec{}\right)$ & & Total & SDSS & SDSS+DES & LSST \\
\tableline
$7$ & $1.0$ & $23.5$ & $45\pm7$ & $26\pm5$ & $5^{+4}_{-3}$ & $8\pm4$ & $8^{+5}_{-3}$ \\
$7$ & $1.0$ & $24.0$ & $45\pm7$ & $24\pm5$ & $4^{+3}_{-2}$ & $7^{+3}_{-4}$ & $7\pm3$ \\
$7$ & $1.5$ & $23.5$ & $45\pm7$ & $13\pm4$ & $2^{+3}_{-1}$ & $4^{+3}_{-2}$ & $4^{+3}_{-2}$ \\
$7$ & $1.5$ & $24.0$ & $45\pm7$ & $10\pm3$ & $2^{+2}_{-1}$ & $3^{+3}_{-2}$ & $3\pm2$ \\
\tableline
$8$ & $1.0$ & $23.5$ & \fieldgalaxyresult{} & $29\pm5$ & $6^{+4}_{-3}$ & $9\pm4$ & $9^{+5}_{-4}$ \\
$8$ & $1.0$ & $24.0$ & \fieldgalaxyresult{} & $27\pm5$ & $5\pm3$ & $7^{+4}_{-3}$ & $8^{+3}_{-4}$ \\
$8$ & $1.5$ & $23.5$ & \fieldgalaxyresult{} & $15\pm4$ & $3\pm2$ & $4^{+3}_{-2}$ & $4^{+3}_{-2}$ \\
$\mathbf{8}$ & $\mathbf{1.5}$ & $\mathbf{24.0}$ & $\mathbf{52\pm7}$ & $\mathbf{12\pm3}$ & $\mathbf{2^{+2}_{-1}}$ & $\mathbf{3^{+3}_{-2}}$ & $\mathbf{4\pm2}$ \\
\tableline
$9$ & $1.0$ & $23.5$ & $58\pm8$ & $33\pm6$ & $6^{+5}_{-3}$ & $9^{+5}_{-4}$ & $10^{+5}_{-4}$ \\
$9$ & $1.0$ & $24.0$ & $58\pm8$ & $30\pm5$ & $5\pm3$ & $8\pm4$ & $8^{+5}_{-3}$ \\
$9$ & $1.5$ & $23.5$ & $58\pm8$ & $16\pm4$ & $3\pm2$ & $4^{+4}_{-2}$ & $5^{+2}_{-3}$ \\
$9$ & $1.5$ & $24.0$ & $58\pm8$ & $13\pm4$ & $2^{+2}_{-1}$ & $3^{+3}_{-2}$ & $4^{+3}_{-2}$ \\
\tableline{}
	\end{tabularx}
\end{table*}
\section{Discussion and Conclusions}%
\label{sec:Conclusions}%
In this \textit{Letter}, we provide quantitative predictions of the size and luminosity function of the population of isolated UDGs within \Mpc[2.5] of the Local Group (which we call ``Local Group field UDGs''), and estimate how many could be detectable in dedicated searches of current data sets, and in future surveys. We produce these predictions using the populations of UDGs in the highest-resolution hydrodynamic simulations from the \Hestia{} suite that are constrained to reproduce the local large-scale structure at \z[0]. This is the first time that such spatially extended galaxies have been simulated self-consistently in such environments (see \figrefs{fig:Methods:re_vs_mu}{fig:Methods:LG_projection_UDGs_marked}). To obtain our results, we rescale the simulations to a common mass, ${\MTot[{\Mpc[2.5]}]=\targetlgmass{}}$, which is consistent with current estimates of \MTot[{\Mpc[1]}] from the Local Group timing argument (see \secref{sec:Results:Mock_sdss_LF}). We predict that there are \maincombinedresult{}~(\percent{68} CL) low-surface-brightness UDGs in the field of the Local Group with stellar masses, ${10^6 \leq \Mstar{}\, /\, \Msun{} < 10^9},$ and effective radii, $\reff{}\geq\kpc[1.5]$; and as many as $27\pm5$ when selecting UDGs with $\reff\geq\kpc[1].$ The UDGs account for approximately one-quarter and one-half, respectively, of the total population of \fieldgalaxyresult{}~(\percent{68} CL) field galaxies with similar stellar masses. As many as \percent{80} of these systems are within \Mpc[1.5] of the Milky Way--M31 midpoint and cluster close to these two primary haloes (see \figref{fig:Results:Cumulative_distribution}), in agreement with the results of \citet{fattahi_missing_2020}.

All of the UDGs are as bright as the ``classical'' satellite galaxies of the Milky Way (i.e. they are brighter than \MV[-8]; see \figref{fig:Results:combined_abs_app_vband_lf}); however, they are much more spatially extended and have ${\reff{} \geq \kpc[1.5]}$. Therefore they are very diffuse, and have faint effective surface brightnesses that make them difficult to detect against the foreground of Galactic stars and the background of distant galaxies. In the surveys that we consider, we find that the detectability of field UDGs is limited most strongly by their faint effective surface brightness; however, we also find that some field UDGs could be detectable in existing survey data sets and are awaiting discovery by dedicated follow-up searches of archival data (see \figref{fig:Results:combined_abs_app_vband_lf}). To estimate how many could be detectable, we generate mock \SDSS{} observations of the field UDG populations in the three highest-resolution \Hestia{} simulations using survey response functions extrapolated from those described in \citet{koposov_luminosity_2008}. Using these, we predict that there are \NudgsInHestia{} UDGs detectable in the \SDSS{} footprint~(see \figref{fig:Results:mock_sdss_lfs}). Almost half of the UDGs that are detected in each mock observation are misclassified as non-UDGs because of projection effects. The total number of UDGs that are detectable is also subject to the variation in the masses of the three Local Group volumes we use. When renomalizing these to $\MTot[{\Mpc[2.5]}]=\targetlgmass{}$, we find \maincombinedresult{} field UDGs within \Mpc[2.5] of the Milky Way--M31 midpoint, of which \maincombinedsdssresult{} are detectable in the footprint of the \SDSS{} (see \secref{sec:Results:Mock_sdss_LF} and \tabref{tab:Results:NUDGs_LG_mass_sel_criteria}). A full-sky survey with a response function similar to that of the \SDSS{}, \DES{}, or LSST will detect the entire population of field UDGs.

To generate mock \SDSS{} observations, we used a simple model of the \citet{koposov_luminosity_2008} \SDSS{} response function. This depends on several physical properties of the galaxies such as their sizes and luminosities, their orientation with respect to the observer, and their physical locations, i.e. their heliocentric distances and projected positions on the sky. The latter are important because galaxies that are partially or totally obscured by the Milky Way can be more difficult to detect against the high-density Galactic stellar foreground, i.e. the ZoA. In \Hestia{}, we find that \percent{{13^{+20}_{-9}}} (\percent{68} CL) of the total UDG population is in the ZoA on average. We do not account for this when estimating the detection efficiencies of the UDGs, and we further assume that the UDGs do not suffer from dust extinction. Correcting for both of these effects would likely reduce the predicted number of UDGs that are detectable in the surveys.

As we have shown, UDGs are challenging to observe because they are extremely diffuse. However, those that contain large reservoirs of neutral hydrogen, such as most isolated observed UDGs as well as the simulated field UDGs in \Hestia{} (S. Cardona--Barrero et al. 2023, in preparation), could be detected more easily. In \HI{} surveys the neutral hydrogen could appear as ultracompact high-velocity clouds~\citep[UCHVCs;][]{giovanelli_are_2009,adams_catalog_2013}. Recent searches for UCHVCs and other \HI{}-bright systems using ALFALFA \citep[e.g.][]{janesh_five_2019}, DES \citep{tanoglidis_shadows_2021}, and HIPASS \citep{zhou_hipass_2022} have produced promising results that could expand the catalog of targets for dedicated follow-up studies. Our results suggest that there is a population of low-surface-brightness, spatially extended galaxies in the Local Group awaiting discovery.
%

\section*{Acknowledgments}
The authors thank the anonymous referee for a thoughtful report that improved the manuscript.
We also thank Vasily Belokurov, Christopher Conselice, Stefan Gottl\"{o}ber, and Sergey Pilipenko for useful comments on the draft manuscript, and Steven Gillman for helpful discussions.
ON and NIL acknowledge  support from the Project IDEXLYON at  the University of Lyon under the Investments for the Future Program (ANR-16-IDEX-0005) and supplementary  support from La R\'{e}gion Auvergne-Rh\^{o}ne-Alpes. ON is also supported by the Polish National Science Centre under grant 2020/39/B/ST9/03494.
ADC is supported by a Junior Leader fellowship from `La Caixa' Foundation (ID 100010434), code  LCF/BQ/PR20/11770010.
SCB is supported by the Spanish MINECO under grant SEV-2015-0548-18-3.
YH has been partially supported by the Israel Science Foundation grant ISF 1358/18.
JS acknowledges support from the ANR LOCALIZATION project, grant ANR-21-CE31-0019 of the French Agence Nationale de la Recherche.
AK is supported by the Ministerio de Ciencia e Innovaci\'{o}n (MICINN) under research grant PID2021-122603NB-C21.
ET acknowledges support by ETAg grant PRG1006 and by the EU through the ERDF CoE grant TK133.
The authors acknowledge the Gauss Centre for Supercomputing e.V. (\url{www.gauss-centre.eu}) for providing computing time on the GCS Supercomputer SuperMUC-NG in support of the \Hestia{} project.

%



\software{\astropy{} \citep{the_astropy_collaboration_astropy_2018},
          \matplotlib{} \citep{hunter_matplotlib_2007},
          \numpy{} \citep{harris_array_2020},
          \pynbody{} \citep{pontzen_pynbody_2013},
          \python{} \citep{van_rossum_python_2009},
          \scipy{} \citep{virtanen_scipy_2020},
          and
          NASA's Astrophysics Data System.
          }
\section*{Data Availability}%
A repository of reduced data and scripts to produce the figures in this \textit{Letter} is available on GitHub\footnote{Supplementary materials: \github{Musical-Neutron/lg_field_udgs}} and archived in Zenodo \citep{newton_hestia_2023}. Requests for access to the raw \Hestia{} simulation data should be directed to a CLUES Collaboration PI.

\bibliography{archive}{}
\bibliographystyle{mnras}



\end{document}